# Fractional Spin fluctuations and quantum liquid signature in $Gd_2ZnIrO_6$

Birender Singh[1#], Deepu Kumar[1], Vivek Kumar[1], Michael Vogl[2], Sabine Wurmehl[2,3], Saicharan Aswartham[2], Bernd Büchner[2,3] and Pradeep Kumar[1*]

[1]*School of Basic Sciences, Indian Institute of Technology Mandi, Mandi-175005, India*

[2]*Leibniz-Institute for Solid-State and Materials Research, IFW-Dresden, 01069 Dresden, Germany*

[3]*Institute of Solid-State Physics, TU Dresden, 01069 Dresden, Germany*

**Abstract:**

Hitherto, the discrete identification of quantum spin liquid phase, holy grail of condensed matter physics, remains a challenging task experimentally. However, the precursor of quantum spin liquid state may reflect in the spin dynamics even in the paramagnetic phase over a wide temperature range as conjectured theoretically. Here we report comprehensive inelastic light (Raman) scattering measurements on the Ir based double perovskite, $Gd_2ZnIrO_6$, as a function of different incident photon energies and polarization in a broad temperature range. Our results evidenced the spin fractionalization within the paramagnetic phase reflected in the emergence of a polarization independent quasi-elastic peak at low energies with lowering temperature. Also, the fluctuating scattering amplitude measured via dynamic Raman susceptibility increases with lowering temperature and decreases mildly upon entering into long-range magnetic ordering phase, below 23 K, suggesting the magnetic origin of these quantum fluctuations. This anomalous scattering response is thus indicative of fluctuating fractional spin evincing the proximate quantum spin liquid phase in a three-dimensional double perovskite system.

*#email id: birender.physics5390@gmail.com*
*\*email id: pkumar@iitmandi.ac.in*

Quantum spin liquid (QSL), a state sans long-range magnetic ordering owing to strong quantum fluctuations but favors entanglement of spins even over a long-range, has been very fascinating since it was proposed in the 1970s [1]. QSL state, a holy grail which so far has been elusive, is most sought after by the contemporary scientific community. A large number of systems have been proposed to be potential host for these states despite some having long-range ordering at low temperature [2-4] with arguments that precursor of QSL state may emerge even in the paramagnetic phase over a wide temperature range owing to the dynamic quantum fluctuations associated with a complex interplay of spin and orbital degrees of freedom (DoF) [5-8]. Recently, the search has been extended to iridium-based oxides, and quite interestingly even to the rare earth systems [9-15]. In iridium-based oxides the focus is on double perovskites (DP), such as $A_2BIrO_6$, (A = La, Nd, Sm, Gd; B = Zn, Cu, Mg), though these systems do show long-range magnetic ordering at low temperature. The existence of QSL state may be inferred indirectly from light scattering experiments via the observation of a polarization invariant broad continuum instead of sharp magnetic modes, characteristic of a long-range ordered phase [7, 9-11, 16].

Iridium-based DP systems have an intertwined spin, orbital and lattice DoF along with coupled crystalline electric field, and the physics of $Ir^{4+}$ ($5d^5$) DP system is believed to be driven by spin-orbital entangled $J_{eff}=1/2$ state [17-19]. Interestingly, in these systems the relativistic spin-orbit coupled $J_{eff}=1/2$ iridium moments reside on the 3D geometrically frustrated face-centered cubic lattice providing unique symmetry allowed anisotropic interactions leading to the Kitaev type interactions suggesting these DP as potential QSL candidate [13-14]. The effective Hamiltonian with multiple interactions is given as $H = \sum_{i,j} J_k S_i^\alpha S_j^\alpha + J\vec{S}_i.\vec{S}_j$ [13], where $J_k$ and $J$ are the Kitaev and Heisenberg parameters, respectively; $\alpha$ is component of the spin directed perpendicularly to the bond connecting spins $(i, j)$, and the ground state, $|GS\rangle$, with this model Hamiltonian is found to be A-type antiferromagnet in line with the experiments [13]. In case of 5$d$ system, the strong spin-orbit coupling (SOC) is expected to quench the orbital DoF, and only spin channel are considered to be active. However, it has been advocated that even in case of 5$d$ system the Jahn-Teller mechanism does effect $t_{2g}$ orbitals in spite of the strong SOC [20-21], thus suggesting non-zero contribution of orbital DoF in controlling the physics of these system. In fact, a spin-orbital entangled quantum spin liquid state is reported in another iridium-based system suggesting the crucial role of these DoF [22]. In addition to spin-orbital entangled mechanism, recently it has been shown that in the three-dimensional (3D) geometrically frustrated systems, the QSL state may be induced due to formation of random-singlet state accompanied with static Jahn-Teller distortions [23]. The system under study, DP Gd$_2$ZnIrO$_6$, undergo a magnetic transition at ~ 23 K ($T_N$) with a canted antiferromagnetic ordering attributed to the interplay between Gd and Ir magnetism, and the transport measurements also evinced the signature of incoherent spin fluctuations [24]. The crystalline electric field ground state ($^8S_{7/2}$) of Gd$^{3+}$ is composed of Kramer's doublets which allows quantum tunneling owing to sizable component of $|J, J_z\rangle$ with a small $|J_z\rangle$ and is well separated from the first excited state, therefore providing a way to realize the QSL state [15, 25, 26] in these rare-earth system as suggested recently.

Motivated by these concrete suggestions for a possible QSL state in these iridium-based DP systems, we undertook an in-depth inelastic light (Raman) scattering studies to probe the quasiparticle excitations in Gd$_2$ZnIrO$_6$, with Ir$^{4+}$ (5$d^5$). Smoking gun evidence for a QSL phase may be uncovered via the observation of a quantum spin and/or orbital fluctuations, and these dynamic fluctuations may reveal itself indirectly via interacting with the photon which is inelastically scattered by these underlying quantum fluctuations. Raman scattering is a very powerful technique to probe the dynamical quantum fluctuations associated with spin and orbital DoF reflected via the emergence of the quasi-elastic peak at low energy in the Raman response $\chi''(\omega,T)$ [27-33], smoking gun evidence of a quantum spin liquid state. Within this scenario, an inelastic light (Raman) scattering can be a suitable technique for the detection of robust signature of a QSL state. The unique ability of Raman scattering in the present study is reflected in the observation of a strong quasi-elastic response with lowering temperature, quite startling it start emerging much above the long-range magnetic ordering temperature. The low energy quasi-elastic peak is found to be nearly isotropic with respect to the polarization of light fixed in the basal plane and is consistent with the constraints imposed by the symmetry. This characteristic low energy scattering response clearly evince the presence of strong spin-orbital coupled underlying quantum fluctuations. Quite interestingly, the corresponding estimated dynamic Raman susceptibility, $\chi^{dyn}(T)$, amplitude does not quenched below $T_N$, instead it decreases only by ~ 25 % of its maximum value, as expected for a conventional magnetic system, signaling that it emerges from a quantum liquid state. This also hint that the ordered ground state in this system may be proximate to quantum phase transition into a spin liquid ground state. Raman scattering measurements on $Gd_2ZnIrO_6$ polycrystalline samples, synthesized as described in ref. 24, were performed in quasi-back scattering configuration using 532-nm and 633-nm Laser at very low power, in the temperature range of 4-330 K [34-35]. To the best of our knowledge, hitherto there is no report on this DP system probing underlying quantum spin fluctuations and exploring the link with the anticipated spin fractionalization a necessary denominator of the quantum liquid $|GS\rangle$. The fluctuation-dissipation theorem connects the underlying dynamic fluctuations to the imaginary part of the corresponding susceptibility which is reflected in the Raman scattering.

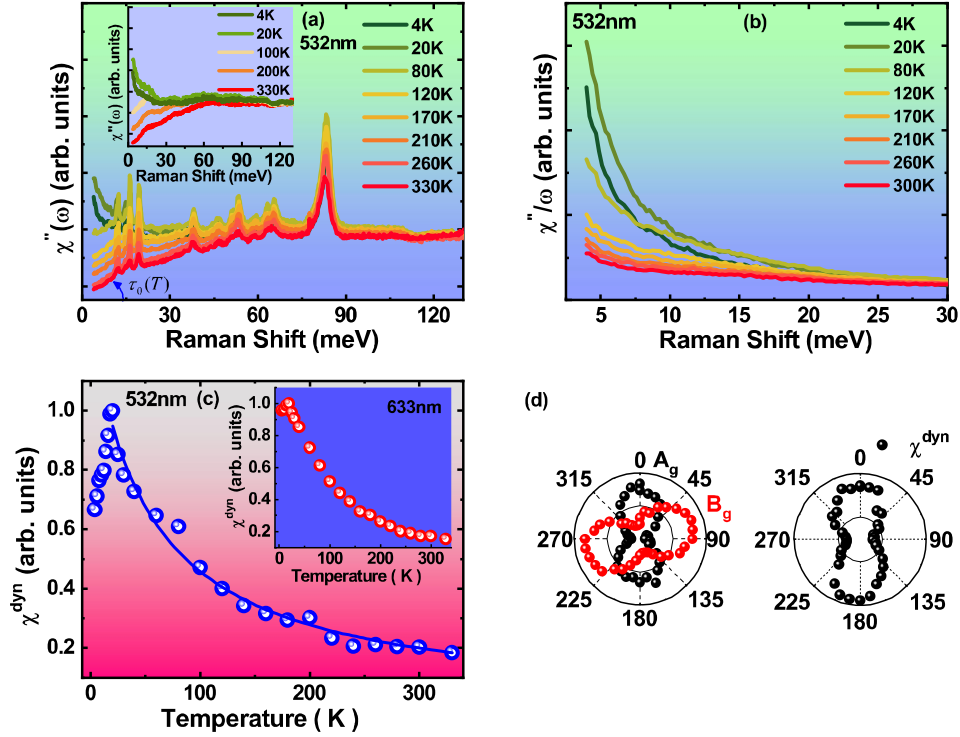

**FIGURE 1:** (Color online) **(a)** Temperature evolution of the Raman response $\chi''(\omega,T)$ (measured raw Raman intensity/$1+n(\omega)$]). Inset shows the phonons subtracted Raman response at selected temperature. **(b)** Temperature dependence of the phonons subtracted Raman conductivity $\chi''(\omega,T)/\omega$. **(c)** Temperature dependence of the dynamic Raman susceptibility $\chi^{dyn}$, extracted using the Kramer-Kroning relation $\chi^{dyn}(q=0,T)=\frac{2}{\pi}\int_0^\Omega \frac{\chi''(\omega,T)}{\omega}d\omega$; by integrating the finite frequency responses up to $\Omega=30$ meV). Inset shows the dynamic Raman susceptibility at different excitation energy i.e., 1.96 eV (633-nm). The solid line above $T_N$ is a Curie-Weiss fit to the Raman dynamic susceptibility as described in the text. **(d)** Polarization dependence of the integrated intensity of the prominent phonon mode (at ~ 84 meV having $A_g$ symmetry, and ~ 54 meV having $B_g$ symmetry), and the dynamic Raman susceptibility.

Figure 1(a) shows the imaginary part of the susceptibility (i.e., Raman response, which reflects the dynamic properties of the collective underlying excitations) $\chi''(\omega,T)$, obtained from the observed Raman intensity $[\propto [1+n(\omega)]\chi''(\omega,T)]$ by dividing it by the Bose factor $[1+n(\omega)]$, as a function of frequency (Raman shift) for different temperature. It is clear from the spectra that the narrow phonon modes are superimposed on the slowly varying continuum. Systematic analysis of this continuum may provide the crucial information about the underlying long wavelength dynamical spin and/or orbital fluctuations via the dynamic Raman susceptibility. The Raman response shows a significant buildup of the intensity below ~ 800 cm$^{-1}$ (100 meV) on lowering the temperature [see Fig. 1(a) and its inset], and quite surprisingly it decreases only slightly upon entering into the spin solid phase. This characteristic scattering feature is typical of the scattering from underlying quantum spin fluctuations. To make a quantitative estimate, here, we focus on the dynamic Raman susceptibility ($\chi^{dyn}(q=0,T) = \lim_{\omega \to 0} \chi(\omega, q=0, T)$) which is obtained from the Raman response at finite frequency using the Kramers-Kronig relation [36] as:

$$\chi^{dyn}(q=0,T) = \frac{2}{\pi}\int_0^\Omega \frac{\chi''(\omega,T)}{\omega}d\omega \qquad \text{------ (1)}$$

The dynamic Raman susceptibilities obtained by integrating the finite frequency responses up to $\Omega$ (= 30 meV) after subtracting the phonon response, the upper cutoff value is chosen as 30 meV, where Raman conductivity $\chi''(\omega,T)/\omega$ [see Fig. 1(b)] shows no change with further increase in the energy, are shown as a function of temperature in Fig. 1(c). With decreasing temperature $\chi^{dyn}$ shows nearly temperature independent behavior down to ~240-250 K, and upon further cooling it keeps increasing down to ~ 25 K, and below 25 K it decreases mildly till 4 K. As in the paramagnetic phase, spins are expected to have no correlation, and $\chi^{dyn}$ should be temperature independent. The significant build of $\chi^{dyn}$ in the temperature range of ~250 K to ~25 K demonstrate the finite entanglement of spin and orbital DoF in the spin gas phase, hinting that this characteristic temperature (~250 K) corresponds to crossover from a conventional paramagnet to the proximate quantum liquid state. Our results clearly suggest the strong growth of the dynamic quantum fluctuations associated with spin and orbital DoF in the paramagnetic phase, which are only mildly quenched below $T_N$ as opposed to a conventional magnetic system, where fluctuations quenched to zero quickly below $T_N$ in the ordered phase [37] and decay exponentially fast above $T_N$. In systems with quantum liquid state the dynamic correlation function may show peculiar temperature and frequency dependence even below the temperature where static correlations saturate [5-7]. Recently, it was advocated that the dynamical spin fluctuations in the paramagnetic state are strongly influenced by the fractionalization of quantum spins [5], therefore the signature of quantum spin fractionalization may be visible in the dynamical measurable properties such as Raman susceptibility. Specifically, it was shown that the dynamical structure factor, which is related to the spin correlation factor exhibits emergence of a low energy quasi-elastic response at low temperature, and is postulated as smoking gun evidence for the fractionalized spins in QSL state. The emergence of a quasi-elastic peak at low energy in our measurements evince the presence of fractionalized spins deep into the paramagnetic phase signaling the proximate QSL state. This anomalous temperature evolution and the distinctive symmetry (to be discussed later) of this low frequency quasi-elastic peak cannot be captured by the conventional long-range magnetic scattering, rather it evinces its intimate links to the underlying spin-orbital entangled quantum liquid phase, seemingly consistent with the theoretical predictions for a QSL state. Temperature evolution of $\chi^{dyn}(T)$ [see Fig. 1(c)], above $T_N$ is fitted well using a canonical Curie-Weiss law of the form $\chi^{dyn}(T) = \frac{\Delta}{T-T_0}$, where $\Delta$ is a constant, and the estimated absolute Curie-Weiss temperature $T_0$ is ~ 55 K and is quite higher than $T_N$ suggesting that it is free from the effects of the long-range magnetic ordering.

We also did our measurements with different incident photon energy [see inset of Fig. 1(c)], the temperature dependence of dynamic Raman susceptibility, $\chi^{dyn}(T)$, is found to be independent of the incident photon energy. This invariance of $\chi^{dyn}(T)$ with respect to the incident photon energy suggests that the resonant terms in the Raman vertex does not modify the temperature behavior of the dynamic susceptibility response, and support the use of the effective mass approximation [32, 36]. In a system where both spin and orbital DoF are entangled, as in the present case, then both DoF are expected to contribute to the total Raman response of the system. In such cases, the total Raman response $\chi''(\omega,T)$ may be given as sum of contribution from both these DoF i.e., $\chi''(\omega,T) = \chi''_{spin}(\omega,T) + \chi''_{orb./electr.}(\omega,T)$. The $\chi''(\omega,T)$ may be given as the imaginary part of the correlation function of magnetic Raman tensor and stress tensor given as [33, 36, 38-40]:

$$\chi(q,\omega) = -i\left[\int_0^\infty dt\, e^{i\omega t} \langle[\tau^+_{\alpha\beta}(q,t), \tau_{\alpha\beta}(0,0)]\rangle + \int_0^\infty dt\, e^{i\omega t} \langle[T^+_{\alpha\beta}(q,t), T_{\alpha\beta}(q,0)]\rangle\right] \quad \text{------ (2)}$$

where $\tau_{\alpha\beta}(r, q \to 0) = \sum_\mu K_{\alpha\beta\mu}(r)S_r^\mu + \sum_{\mu,\nu} G_{\alpha\beta\mu\nu}(r)S_r^\mu S_r^\nu + \sum_\delta \sum_{\mu,\nu} H_{\alpha\beta\mu\nu}(r,\delta)S_r^\mu S_{r+\delta}^\nu + --$, first term leads to the scattering from single spin fluctuations, second and third term gives rise to the scattering by pairs of spin fluctuations. Tensor $K$, $G$ and $H$ describes the strength of the coupling between the incident light and underlying magnetic DoF. $T_{\alpha\beta}(q) = \sum_{q,\sigma,\alpha,\beta} \gamma_{\alpha\beta}(q) C^+_{q\alpha\sigma} C_{q\beta\sigma}$ is the stress tensor, and Raman vertex, $\gamma_{\alpha\beta}(q)$, within the effective mass approximation, valid in non-resonant case as here, is given as $\gamma_{\alpha\beta}(q \to 0) = \frac{1}{\hbar^2} \sum_{r,s} e_r^i \frac{\partial^2 \varepsilon_k^{\alpha\beta}}{\partial k_r \partial k_s} e_s^f$, where $e^i$ and $e^f$ are the polarization vector of the incident and scattered light, respectively. For our experimental geometry, the incident and scattered light is confined within the XY plane, we have $e^i = \frac{1}{\sqrt{2}}(\hat{x}+\hat{y}) = e^s$; for this symmetry configuration one expect Raman vertex, $\gamma_{\alpha\beta}(q)$, to be same for both $A_g$ and $B_g$ symmetry i.e., orbital fluctuations may be seen in both these channel, hence the Raman response will be invariant, and is consistent with our experiments [see Fig. 1(d) for $\chi^{dyn}$ response]. Also, in a recent report it was suggested that the orbital fluctuations may be observed in both $A_{1g}$ and $B_{1g}$ channel for $D_{4h}$ point group [29], which translates to $A_g$ and $B_g$ for $C_{2h}$ point group in the present case. Due to the lack of local order parameter a very weak or no polarization dependence magnetic Raman response is advocated to be one of the key signatures of QSL state [41-42]. The nearly isotropic behavior of Raman response [see Fig. 1(d) for $\chi^{dyn}$ response] agree very well with the suggestion based on theoretical calculations. We note that, there is one caveat in our polarization dependent response i.e., we have used the polycrystals for our measurements. Probably, the single crystal studies may be more detailed; however, we note that the phonon modes do show correct polarization dependence similar to our earlier polarization measurements on single crystal of $Nd_2ZnIrO_6$ [19], evidencing that nearly polarization invariant behavior of the continuum is intrinsic to the system. Now, we discuss the frequency and temperature dependence of this polarization invariant low-frequency response. As is clear from the spectra [see Fig. 1(a)], this fluctuations response may be decomposed into two contributions, a quasi-elastic peak in the low frequency regime (below ~ 200 - 300 cm$^{-1}$), and a broad continuum. To quantify the temperature dependences of these two components contributing to the dynamic fluctuations, we fit the data using the following general expression [27, 30]: $\chi''(\omega,T) = \chi''_{QEP}(\omega,T) + \chi''_b(\omega,T)$, where the first part, i.e., quasi-elastic peak (QEP) is modeled by a damped Lorentzian:

$$\chi''_{QEP}(\omega,T) = A(T)\frac{\Gamma(T)\omega}{\omega^2 + \Gamma^2(T)} \qquad \text{------ (3)}$$

where $A$ is the quasi-elastic scattering amplitude, and $\Gamma$ is the linewidth of the peak reflecting the fluctuation rate. And the broad continuum, $\chi''_b(\omega,T)$, is fit using a third-order polynomial with only odd powers in $\omega$ to guarantee causality i.e., $\chi''_b(\omega,T) = B_1(T)\omega + B_2(T)\omega^3$.

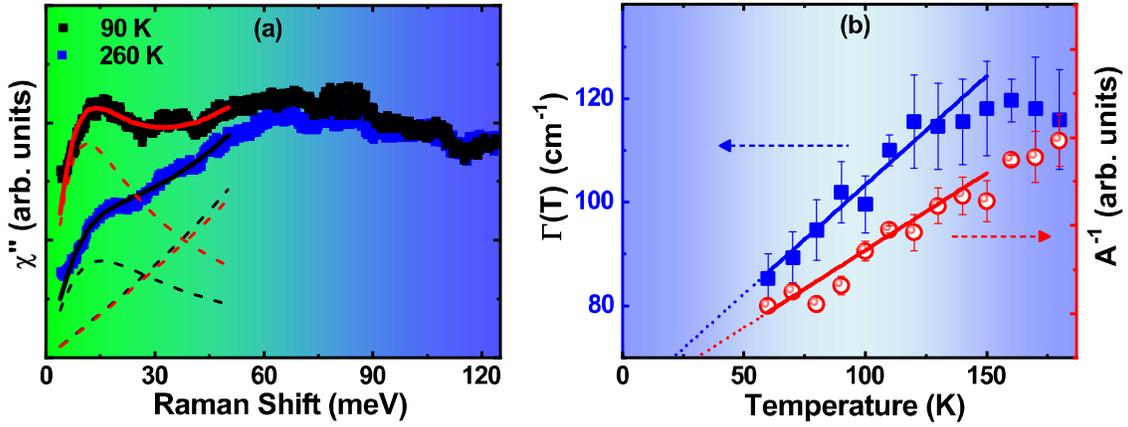

**FIGURE 2:** (Color online) **(a)** Low energy fits using a damped Lorentzian, and an odd in frequency polynomial to the Raman susceptibility [i.e., $\chi''(\omega,T) = \chi''_{QEP}(\omega,T) + \chi''_b(\omega,T)$; where $\chi''_{QEP}(\omega,T) = A(T)\frac{\Gamma(T)\omega}{\omega^2 + \Gamma^2(T)}$ and $\chi''_b(\omega,T) = B_1(T)\omega + B_2(T)\omega^3$] at two different temperature. **(b)** Temperature dependence of the inverse of the area $A(T)$ and linewidth ($\Gamma(T)$) of the quasi-elastic peak. The solid lines are the linear fits between 60 K and 130 K.

As is clear from Fig. 2(a) that above equation fits well the Raman response data at low energy, up to ~ 400 - 500 cm$^{-1}$. Figure 2(b) shows the temperature dependence of the inverse of $A$ and linewidth, $\Gamma$, of the QEP. The characteristics of this QEP may be linked to the mass, M(T), in the fluctuation propagator [36, 43]. At temperatures where quantum fluctuations dominate, the mass M(T), inverse of spectral weight $A(T)^{-1}$ and the fluctuation rate, $\Gamma(T)$, are linear in $T$ i.e., $A(T)^{-1} \propto (T-T^*)$ and $\Gamma(T) \propto (T-T^{**})$. In the ordered state M(T) saturate, hence the extrapolation to zero of the linear part of $A(T)^{-1}$ may provide a good estimate of $T^*$. Following this, the temperature dependences of $A(T)^{-1}$ and $\Gamma(T)$ were fitted between 60 and 130 K using a linear form $A(T)^{-1} = a(T-T^*)$ and $\Gamma(T) = g(T-T^{**})$ [see solid lines in Fig. 2(b)]. Experimentally, the absolute $T^*$ is found to be 55 K similar to $T_0$, as obtained from the canonical Curie-Weiss fit of the dynamic Raman susceptibility, $\chi^{dyn}(T)$. However, the zero-temperature intercept of the QEP linewidth, $\Gamma$, absolute $T^{**}$ is significantly higher (145 K). One expect that both these temperature be in similar range, the difference between $T^*$ and $T^{**}$ may be understood by the temperature dependence of the single particle scattering rate $g$, which may be gauged via transport measurements. We note that the low energy QEP is well reproduced by a damped Lorentzian, especially above ~ 60 K, $\chi''_{QEP}(\omega,T) = A(T)\frac{\Gamma(T)\omega}{\omega^2+\Gamma^2(T)}$. $A(T)^{-1}$ extrapolating to zero close to $T_0$, and the fluctuation rate, $\Gamma(T)$, which is linked to the relaxational dynamics of the fluctuations of QEP shows a softening with decreasing temperature [see Fig. 2(b)]. The temperature evolution of $A(T)$ matches qualitatively with the dynamic Raman susceptibility $\chi^{dyn}(T)$, suggesting that the temperature dependent behavior is dominated by this low frequency QEP.

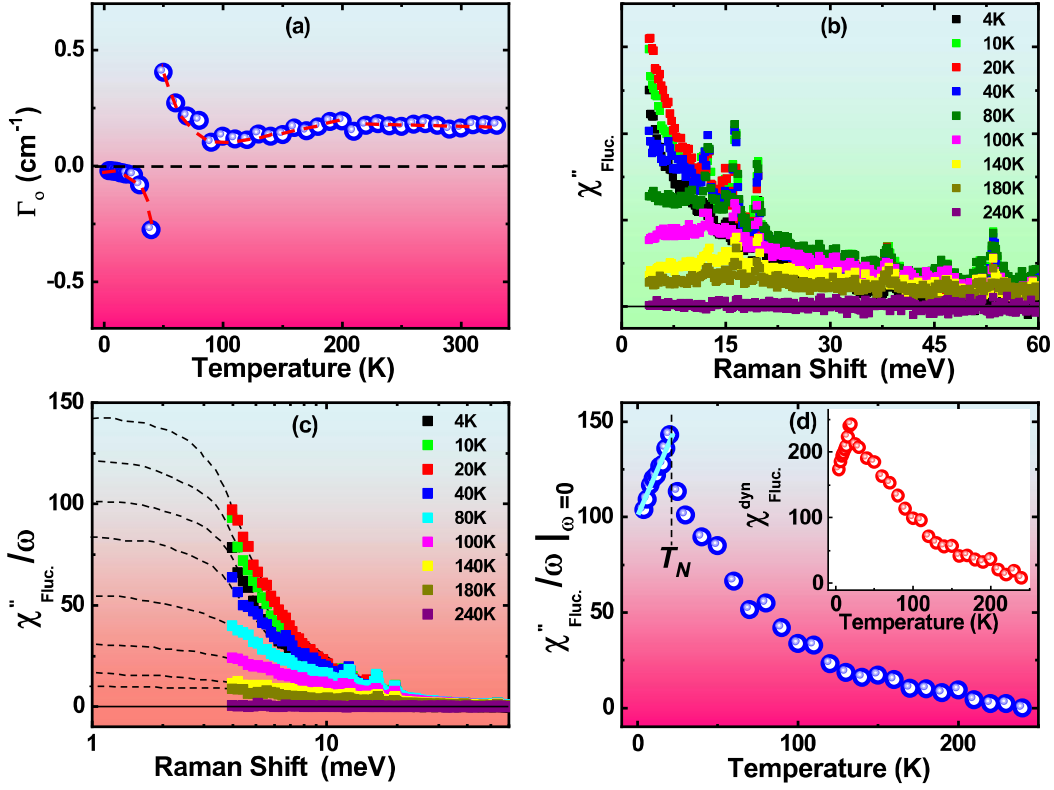

**FIGURE 3:** (Color online) **(a)** Raman relaxation rates $\Gamma_0(T)$ [$= \left( \dfrac{\partial \chi''(\omega,T)}{\partial \omega} \right)^{-1} \Big|_{\omega=0}$] as a function of temperature. Dash red line and horizontal black line are guide to the eye. **(b)** Fluctuation contribution to the Raman response extracted from full Raman response by carefully subtracting the response at 250 K from all the lower temperature data. **(c)** Temperature dependence of fluctuating Raman conductivity, $\chi''_{Fluc.}(\omega,T)/\omega$ as a function of logarithmic Raman shift on x-axis. **(d)** Temperature dependence of the initial slope of the fluctuating Raman conductivity, as shown in (c), extracted as described in the text. Inset shows the temperature dependence of the fluctuating dynamic Raman susceptibility $\chi^{dyn}_{Fluc.}$, extracted using the Kramer-Kroning relation ($\chi^{dyn}_{Fluc.}(q=0,T) = \dfrac{2}{\pi} \int_0^{\Omega} \dfrac{\chi''_{Fluc.}(\omega,T)}{\omega} d\omega$; by integrating the finite frequency responses up to $\Omega = 30$ meV).

We also analyzed our data from a different perspective i.e., by carefully extracting the fluctuation part, after subtracting the continuum which is almost constant. Interestingly, we found that the temperature dependence of the Raman response is dominated by the fluctuation part which is reflected in the similarity of temperature evolution of the low energy spectrum in different analysis performed earlier. To extract the contribution of fluctuations part, we followed the following procedure. Figure 1(a) shows the temperature evolution of Raman response, $\chi''(\omega,T)$, for Gd$_2$ZnIrO$_6$. The initial slope $\tau_0(T)$ [see Fig. 1(a)], using memory function method, may be related with static transport relaxation rate $\Gamma_0(T)$ [$\equiv \Gamma_0(\omega \rightarrow 0, T)$] as

$$\hbar[\tau_0(T)]^{-1} = \Gamma_0(T) = \left(\frac{\partial \chi''(\omega,T)}{\partial \omega}\right)^{-1}\bigg|_{\omega=0}$$

[44-46]. Figure 3(a) shows that above ~ 220-240 K, $\Gamma_0(T)$ is nearly constant. Below ~ 200 K it starts showing decreasing trend, and is simultaneously accompanied by the increase in the intensity gain of the spectrum below ~ 200-300 cm$^{-1}$ [see Fig. 1(a)]. We note that similar temperature dependence changes are also reflected in the $\chi^{dyn}(\omega,T)$ [see Fig. 1(c)]. This increase in the intensity with decreasing temperature points that an additional contribution other than the continuum is building up in the background and have its origin in the fractional quantum spin fluctuations, and potentially signal the transition from the high temperature paramagnetic phase to the quantum liquid state below ~ 220 K.

Therefore, we have used this temperature as a mark of the crossover temperature to separate the quantum fluctuations response from the nearly constant continuum. For simplicity, we have considered that above ~ 220 K contribution to the Raman response is only from the continuum and subtracted the 250 K data from all the spectra at lower temperatures. The Raman response obtained after subtraction is shown in Fig. 3(b), and it increases rapidly with decreasing temperature without any divergence. Interestingly, the fluctuating Raman response, $\chi''_{Fluc.}(\omega,T)$, does not quenched below $T_N$ as one would expect in a conventional magnetic system owing to the long-range order. Rather, the intensity decreases only slightly below $T_N$. To gauge the quantitative temperature evolution of this fluctuating part, we estimated the dynamic Raman susceptibility from this fluctuation part, $\chi^{dyn}_{Fluc.}(\omega,T)$ [see inset of Fig. 3(d)], and it shows similar behavior as that of $\chi^{dyn}(\omega,T)$ (see Fig. 1(c)), suggesting that the temperature evolution of the Raman response is dominated by the fluctuating part only. Another way of gauging temperature dependence of the intensity of the fluctuating part may be done using initial slope of the fluctuating Raman response [28], which is proportional to the intensity and is expected to follow $1/T-T_0$ behavior within the mean field theory. Using $\chi''_{Fluc.}(\omega,T)$, one may extract the initial slope by plotting Raman conductivity, $\chi''_{Fluc.}(\omega,T)/\omega$, at each temperature, and the temperature dependence of the initial slope may be directly read off from graph by simply plotting Raman conductivity against a logarithmic energy scale and extrapolating the same to zero frequency [see Fig. 3(c)]. We extracted the slope using the procedure described above [see Fig. 3(d)]. Quite surprisingly, we find an excellent agreement with the dynamic Raman susceptibility, $\chi^{dyn}$, (see Fig. 1(c)) extracted above, and qualitatively it also matches quite well with the intensity extracted using $\chi^{dyn}_{Fluc.}(\omega,T)$ (see inset of Fig. 3(d)).

Here, we analyzed the background having a quasi-elastic peak which gains strength with lowering temperature, along with a broad continuum, which is nearly temperature independent, using dynamic Raman response and carefully extracted the fluctuation response. Our results obtained via different analysis depicts consistent anomaly in the vicinity of the same temperature i.e., ~ 200 - 220 K, suggesting that this is the temperature where fractionalization of the spins and onset of proximate quantum liquid phase start. These anomalies also evinced that the underlying continuum arises mainly from the fluctuation of the spin fractionalization. In summary, our in-depth temperature, polarization and different incident photon energy dependent Raman studies demonstrates the signature of spin fractionalization in iridium-based double perovskite $Gd_2ZnIrO_6$. The anomalous emergence of a long wavelength polarization invariant quasi-elastic peak, which reflects the fluctuation of the fractionalized spins, points that these three dimensional geometrically frustrated iridium-based double perovskite systems potentially realizes the quantum liquid state.

**Acknowledgment**

PK thanks IIT Mandi for the experimental facilities and DST India for the financial support. The authors at Dresden thank Deutsche Forschungsgemeinschaft (DFG) for the financial support via Grant No. DFG AS 523/4-1 (S.A.) and via project B01 of SFB 1143 (project-id 247310070).